# Adaptive Self-Optimization of Polarization Efficiency for Low-Power NV Center Thermometers Based on PAS-Q

Yifan Wang[1,2], Li Xing[2*], Ruigang Wang[3], Zheng Qian[1*]

[1]*School of Instrumentation Science and Opto-electronics Engineering, Beihang University, Beijing 100191, China*

[2]*Division of Thermophysics and Process Measurements, National Institute of Metrology, NIM, Beijing 100029, China*

[3] *Hangzhou Normal University*

*Corresponding author: xingli@nim.ac.cn;

**abstract**—Nitrogen-vacancy (NV) centers in diamond have attracted considerable attention for micro- and nanoscale thermometry owing to their exceptional spatial resolution and high measurement precision. High-power or long-duration laser excitation is typically required to achieve sufficient electron spin polarization and a high measurement signal-to-noise ratio (SNR). However, such excitation conditions introduce significant laser-induced heating effects, which severely degrade the accuracy and stability of temperature measurements. To address this fundamental trade-off, a reinforcement learning (RL)-based approach is proposed to optimize polarization efficiency under low-power constraints. First, a comprehensive physical model is established by integrating five-level optical pumping rates with the spin polarization dynamics of NV centers. A polarization constraint model for minimizing heating effect is constructed by characterizing the temperature rise behavior of pulsed optically detected magnetic resonance (P-ODMR) under different laser power and pulse width in an external magnetic field. Subsequently, a RL control framework based on the Progressive Action-Space Q-learning (PAS-Q) algorithm is proposed. The framework dynamically explores the physical parameter space to identify optimal combinations of laser power and pulse width under strict temperature-rise constraint, thereby maximizing spin polarization efficiency. The RL agent is trained using numerical simulations of the physical environment to derive an optimal control strategy. Finally, the strategy is implemented in the hardware system, and the laser heating effect has been reduced by 44%. The spin polarization rate of NV centers have been increased by 33% through analysis of $T_1$ relaxation dynamics, meanwhile the parameter optimization process is 71% faster than that of traditional optimization methods. This work achieves highly efficient electronic spin polarization under low laser power and heating effect, as well as the efficiency of parameter optimization is greatly improved. Moreover, this method can be extended to other quantum spin sensors such as silicon-vacancy centers in diamond and divacancy defects in silicon carbide, which are susceptible to heating effects of applied physical fields.

# I introduction

With the rapid development of chip manufacturing and biomedicine, traditional temperature measurement methods such as thermocouples and infrared thermometry, struggle to simultaneously meet the requirements of measurement accuracy and spatial resolution in special scenarios. Consequently, high-precision temperature measurement at the micro-nanoscale has become a critical scientific issue that urgently needs to be solved. Recently, the development of quantum precision measurement and optical manipulation technologies has created new opportunities to solve this problem. Among them, nitrogen-vacancy (NV) centers in diamond have emerged as a core physical carrier for micro-nanoscale quantum sensing owing to their theoretical capability for ultra-high precision multiparameter measurements, including temperature, magnetic field, and electric field sensing [1], as well as their characteristics of easy quantum manipulation at room temperature, good biocompatibility, and stable physicochemical properties. It is expected to solve the problem of lacking high-precision non-destructive in-situ measurement methods across cutting-edge research fields [2,3].

In order to improve the spin polarization of NV centers and achieve a high measurement SNR, it is necessary to apply high-power or long-duration excitation laser to NV centers. However, a large amount of light energy is dissipated through non-radiative intersystem crossing (ISC) and phonon-scattering channels [4] during this excitation process, leading to thermal accumulation in the local lattice and increasing the uncertainty of temperature measurement[5]. Conversely, excessively low laser power results in insufficient NV-centers polarization, thereby degrading the SNR or requiring prolonged long polarization times [6]. In 2016, S. M. Blakley et al. [7] demonstrated that the heating effect of external fields like excitation laser, can suppress fluorescence intensity emitted by NV centers. In 2020, Scott E. Lillie et al. [8] directly reduced the laser power used in P-ODMR, but this led to a decrease in signal intensity and SNR, which ultimately results in a decrease in temperature measurement accuracy. Therefore, achieving an optimal balance between spin polarization efficiency and laser heating effect has become a research focus in NV centers thermometry, which is essential for its future practical applications. In 2022, Ouyang et al. [9] established a thermal model by measuring the zero-field splitting (ZFS) parameter $D$ under different laser powers in continuous wave (CW) ODMR experiments to compensate for frequency shifts of $D$ values induced by laser heating effect. However, this static power compensation scheme based on CW-ODMR cannot physically block the substantial interference of laser heating effect in thermometry systems, nor can it effectively maintain high spin polarization efficiency under low power conditions. In 2023, Li et al. [10] developed a common-mode rejection (CMR) model to suppress fluctuation noise and heating effect caused by high laser power in NV magnetometers. After experimentally optimizing the laser power and modulation frequency, the magnetic field sensitivity of the system improved from 4.49 nT/Hz$^{1/2}$ to 790.8 pT/Hz$^{1/2}$, and the noise level was reduced by a factor of 6.2. Although this method effectively suppressed the influence of laser interference on NV sensing, it also increased system complexity. In 2024, Hu et al. [11] proposed a real-time frequency tracking method to compensate for resonance frequency drift caused by thermal noise, significantly improving spin-state readout contrast during long-duration experiments. Nevertheless, this method is a passive compensation strategy and cannot suppress heating effect of laser from the source. In 2025, Shi et al. [12] and Leon Advena et al. [13] reduced local temperature rise by adding heat dissipation substrates and controlling the NV centers density. John F. Barry et al. [14] optimized the polarization time and π-pulse time using the method of traditional empirical parameter scans during sensitivity measurements. This approach is not only time-consuming and inefficient but also prone to trapping the system in suboptimal solutions caused by temperature drift. In summary, existing methods have notable limitations: there remains a lack of adaptive control strategies that simultaneously achieve minimal heating, efficient spin polarization, and rapid optimization. In recent years, RL has demonstrated considerable potential in multi-objective optimization [15–18] and has been gradually applied to quantum precision measurement, owing to its efficient optimization capabilities[19]. However, existing RL approaches are not fully compatible with the requirements of full NV polarization and minimal heating. This incompatibility prevents efficient spin polarization in low-power conditions.

We propose an NV spin polarization controller based on the PAS-Q algorithm for efficient optimization of spin polarization under low laser power. Firstly, we construct a polarization constraint model that minimizes heating effect of laser power by integrating the five-level NV dynamics with the temperature-rise characteristics from ODMR measurements. Moreover, an innovative progressive action-space Q-learning control architecture is proposed, which maps spin polarization and thermal relaxation processes to a Markov decision process. It achieves autonomous and efficient global optimization within a two-dimensional parameter space for laser manipulation by introducing segmented thermal safety penalties and dynamic boundary strategies. After large-scale training in a simulated environment, the optimal control strategy is deployed on the hardware platform. Comparative experiments show that the proposed approach reduces optimization time by 71%, decreases heating effect by 44%, and increases the NV centers spin polarization rate by 33%. This method effectively suppresses the laser heating effect from the source, solving the challenges of low polarizability and long polarization time under low laser power.

# Ⅱ. Establishment of NV Centers Polarizability Model

## A. NV Centers Five-Level Energy Model

Both the ground and excited states of the NV centers electronic spin are spin-triplet, corresponding to $m_s$=0, $m_s$=+1, and $m_s$=−1. Since the dynamical behaviors of the $m_s$=+1 and $m_s$=−1 states are nearly identical under laser excitation, the energy level structure of the NV centers can be simplified into a five-level model as illustrated in Fig. 1. In this model, the $m_s$=0 and $m_s$=±1 of the ground spin-triplet state $^3A_2$ are defined as $|1\rangle$and $|2\rangle$ , respectively. The $m_s$=0 and $m_s$=±1 of the excited triplet state $^3E$ are defined as $|3\rangle$and $|4\rangle$ , respectively. Since the transition rate between $^1A$ and $^1E$ is much faster than other dynamical processes, the two singlet states can be approximated as an metastable state $|5\rangle$.

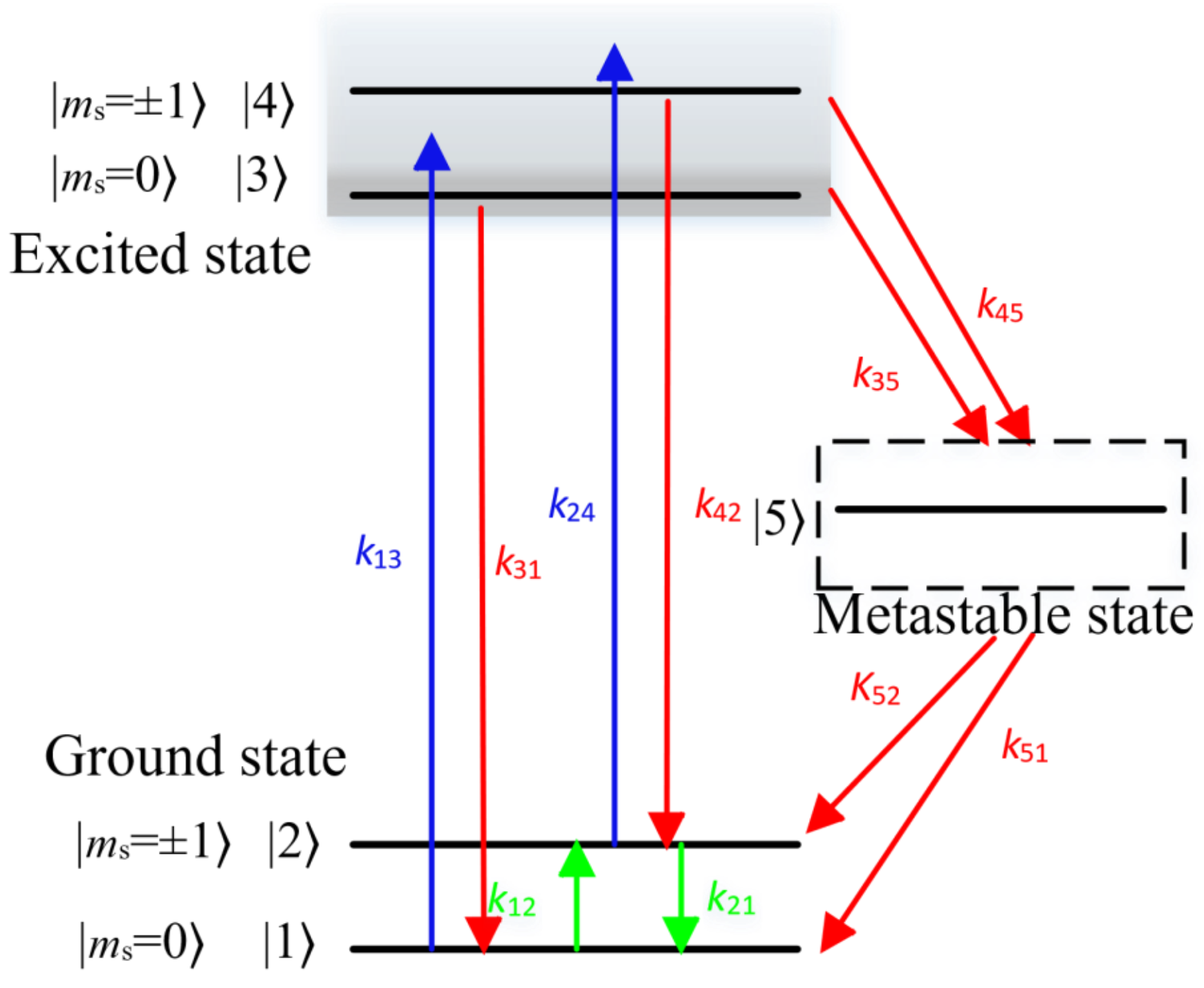

**Fig. 1 Depiction of the NV energy level structure and the corresponding dynamics.**

The difference of transition rates among energy levels during ISC causes a transfer in the population of electronic spin states during laser excitation. The population on energy level $|i\rangle$ is defined as $W_i$, where $i$=1,2,3,4,5. The current state of electronic spins can be represented as the population vector $\vec{W}$,

$$\vec{W} = (W_1, W_2, W_3, W_4, W_5). \tag{1}$$

The total population is normalized such that $\sum_i W_i = 1$. The transition rate from energy level $|i\rangle$ to $|j\rangle$ is defined as $k_{ij}$ ($i, j$ = 1, 2, 3, 4, 5). These rates quantify the net population transfer among energy levels[20], and their definitions are summarized in Table 1.

**TABLE 1 Physical definitions of rate constants.**

| Transition process | Transition rate | Rate description | Note |
|---|---|---|---|
| $|1\rangle$(ground, $m_s$=0) → $|3\rangle$(excited, $m_s$=0) | $k_{13}$ | Laser excitation rate | $k_{13} \propto \alpha P$ ($\alpha$ is the laser power coefficient. $P$ is the laser |

| | | | |
|---|---|---|---|
| | | | power） |
| $\vert 2\rangle$(ground, $m_s$=±1) → $\vert 4\rangle$(excited, $m_s$=±1) | $k_{24}$ | Laser excitation rate | $k_{13}$=$k_{24}$ |
| $\vert 3\rangle$(excited, $m_s$=0) → $\vert 1\rangle$(ground, $m_s$=0) | $k_{31}$ | Radiative decay rate | |
| $\vert 4\rangle$(excited, $m_s$=±1) → $\vert 2\rangle$(ground, $m_s$=±1) | $k_{42}$ | Radiative decay rate | |
| $\vert 4\rangle$(excited, $m_s$=±1) → $\vert 5\rangle$(metastable state) | $k_{45}$ | ISC rate | |
| $\vert 5\rangle$(metastable state) → $\vert 1\rangle$(ground, $m_s$=0) | $k_{51}$ | Non-radiative decay rate | Preferential to $m_s$=0 |
| $\vert 5\rangle$(metastable state) → $\vert 2\rangle$(ground, $m_s$=±1) | $k_{52}$ | Non-radiative decay rate | |
| $\vert 1\rangle$(ground, $m_s$=0) ↔ $\vert 2\rangle$(ground, $m_s$=±1) | $k_{12}$，$k_{21}$ | Longitudinal relaxation rate | $k_{12}$+$k_{21}$=$\gamma$ ($\gamma$ is relaxation rate) |

For example, the energy levels that can transition to energy level $|1\rangle$ are $|3\rangle$, $|2\rangle$, and $|5\rangle$, while $|1\rangle$ can transition to energy levels $|2\rangle$ and $|3\rangle$. Assuming that the population of energy level $|1\rangle$ at time step $n$ is $W_1{}^{(n)}$, the population variation after evolution over a time interval d$t$ can be expressed as follows:

$$\frac{\mathrm{d}W_1}{\mathrm{d}t} = W_5 k_{51} + W_2 k_{21} + W_3 k_{31} - W_1 k_{13} - W_1 k_{12} \,. \tag{2}$$

The rate equations for the five energy levels can be obtained based on the methods in Refs[21,22]. The rate equations that describe the time evolution of the electronic spins in the five-level system can be expressed as:

$$\begin{cases} \frac{dW_1}{dt} = W_5k_{51} + W_2k_{21} + W_3k_{31} - W_1k_{13} - W_1k_{12} \\ \frac{dW_2}{dt} = W_5k_{52} + W_4k_{42} + W_1k_{21} - W_2k_{12} - W_2k_{24} \\ \frac{dW_3}{dt} = W_1k_{13} - W_3k_{31} - W_3k_{35} \\ \frac{dW_4}{dt} = W_2k_{24} - W_4k_{42} - W_4k_{45} \\ \frac{dW_5}{dt} = W_4k_{45} + W_3k_{35} - W_5k_{52} - W_5k_{51} \end{cases} . \quad (3)$$

The rate equations can be transformed into matrix form as follow:

$$\frac{d}{dt}\mathbf{W} = \mathbf{K}\bullet\mathbf{W}, \quad (4)$$

where

$$\mathbf{K} = \begin{bmatrix} -k_{13} - k_{12} & k_{21} & k_{31} & 0 & k_{51} \\ k_{12} & -k_{21} - k_{24} & 0 & k_{42} & k_{52} \\ k_{13} & 0 & -k_{31} - k_{35} & 0 & 0 \\ 0 & k_{24} & 0 & -k_{42} - k_{45} & 0 \\ 0 & 0 & k_{35} & k_{45} & -k_{52} - k_{51} \end{bmatrix}. \quad (5)$$

The lifetimes of the excited state and the metastable state are much shorter than those of the ground state. Therefore, their populations satisfy $\frac{dW_3}{dt} = \frac{dW_4}{dt} = \frac{dW_5}{dt} = 0$, yielding the following equation:

$$\begin{cases} \frac{dW_3}{dt} = W_1k_{13} - W_3k_{31} - W_3k_{35} = 0 \\ \frac{dW_4}{dt} = W_2k_{24} - W_4k_{42} - W_4k_{45} = 0 \\ \frac{dW_5}{dt} = W_4k_{45} + W_3k_{35} - W_5k_{52} - W_5k_{51} = 0 \end{cases} \Rightarrow \begin{cases} W_3 = \frac{W_1k_{13}}{k_{31} + k_{35}} \\ W_4 = \frac{W_2k_{24}}{k_{42} + k_{45}} \\ W_5 = \frac{W_4k_{45} + W_4k_{45}}{k_{51} + k_{52}} = \frac{1}{k_{51} + k_{52}}\left(\frac{k_{35}k_{13}W_1}{k_{31} + k_{35}} + \frac{k_{45}k_{24}W_2}{k_{42} + k_{45}}\right) \end{cases} . \quad (6)$$

The derivation relies on the following physical constraints and assumptions. Firstly, the excitation cross-sections of 532 nm laser for $m_s$=0 (|1⟩→|3⟩) and $m_s$=±1 (|2⟩→|4⟩) are considered approximately equal[23], and the rate $k_{24} = k_{13}$ is set. Secondly, the polarization mechanism relies on the pronounced asymmetry in the ISC rate, i.e. $k_{45} \gg k_{35}$. Thirdly, the contribution of the $m_s$=0 branch to the ISC process is neglected, leading to $k_{35} \approx 0$. The experimentally observed high polarization arises from the preferential decay of the metastable state to the ground state $m_s = 0$, i.e., $k_{51} \gg k_{52}$. The corresponding branching efficiency is defined as $\eta = \frac{k_{51}}{k_{51} + k_{52}}$, and $\eta$ denotes the probability that the metastable state decays into the $m_s = 0$ of ground state. Finally, the ground state populations at thermal equilibrium are denoted by $W_{1\,\mathrm{eq}}$ and $W_{2\,\mathrm{eq}}$, corresponding to the equilibrium populations of states $|1\rangle$ and $|2\rangle$, respectively. $\Gamma_i$ denotes the total decay rate of state $|i\rangle$, defined as the sum of all transition rates originating from that state, i.e., $\Gamma_3 = k_{31} + k_{35}$, $\Gamma_4 = k_{42} + k_{45}$, $\Gamma_5 = k_{51} + k_{52}$. Accordingly, the effective coefficients are given by $C_1 = \frac{k_{51}k_{35}}{\Gamma_3\Gamma_5}$, $C_2 = \frac{k_{51}k_{45}}{\Gamma_4\Gamma_5}$.

The polarization model based on (2) can be given by

$$\frac{\mathrm{d}W_1}{\mathrm{d}t} = -C_1 k_{13} W_1 + C_2 k_{24} W_2 - \gamma W_1 + \gamma W_{1\,\mathrm{eq}}. \tag{7}$$

## B. NV centers polarization rate model

Taking the time derivative of the polarization definition $\rho = \frac{W_1}{W_g}$, yields the following expression:

$$\frac{\mathrm{d}\rho}{\mathrm{d}t} + \left[\alpha P(C_1 + C_2) + \gamma\right]\rho = \alpha P C_2 + \gamma \rho_{\mathrm{eq}}. \tag{8}$$

The transition rate parameters are adopted from Ref. [24], which are summarized in Table 2. The final polarization model can be expressed as follows:

$$\rho(P,t) = W_1^{sat} + (W_1(0) - W_1^{sat}) e^{-\lambda(P)t}, \tag{9}$$

where, $\lambda(P) = \alpha P(C_1 + C_2) + \gamma$, $W_1^{sat} = \frac{\alpha P C_2 + \gamma \rho_{\mathrm{eq}}}{\alpha P(C_1 + C_2) + \gamma}$.

**TABLE 2 Reference values of parameters for the polarization rate model**

| parameters | value | parameters | value |
| --- | --- | --- | --- |
| $k_{31}$ | 82.89 MHz | $k_{52}$ | 3.27 MHZ |
| $k_{35}$ | 3.89 MHz | $\alpha$ | 5.186 MHz/mW |
| $k_{45}$ | 47.10 MHz | $\gamma$ | 1/897.48 $\mathrm{us}^{-1}$ |
| $k_{51}$ | 1.63 MHz | $\rho_{\mathrm{eq}}$ | 1/3 |

The polarization of NV centers electronic spins into the $m_s$=0 state typically relies on optical pumping and ISC process. However, the substantial lifetime difference between the excited state and the metastable state leads to significant spin accumulation in the metastable state under continuous laser excitation[25][26]. This phenomenon not only reduces spin polarization rate but also decreases temperature-measurement SNR owing to the associated heating effect[27]. It is necessary to adjust the laser pulse sequence to overcome this bottleneck. Appropriately shortening each excitation pulse time can interrupt the excitation process before the metastable-state population reaches equilibrium, which promotes rapid spin cycling between the ground and excited states to suppress laser-induced heating effect[28]. Efficient spin polarization can be achieved by introducing an appropriate waiting interval and repeating the optical pumping cycle[29], thereby addressing the insufficient spin polarization caused by a single short pulse[30].

## C. Modeling and parameter measurement of polarization control system

An action-dependent discretized state-space representation is established, which introduce the continuous spin dynamics of NV centers into the RL training framework. This work examines the dynamics of NV centers under pulsed laser excitation, using an analytical-solution-based discretization scheme to ensure numerical stability and physical accuracy at the nanosecond timescale.

The population $\rho_{00}$ of the spin ground state $|0\rangle$ evolves according to a standard rate equation. The system exhibits exponential relaxation toward the steady state $\rho_{ss}$ under a given laser power $P$. At the $k$-th decision step, the agent selects an action by specifying the polarization duration $t$ and the laser power $P$. The mapping that describes the spin-state evolution from the initial thermal state $\rho_{eq}$ = 1/3 to the current observed state can then be expressed as,

$$\rho_{k+1} = \Phi(P_k, t_k) = \rho_{ss}(P_k) + \left[\rho_{eq} - \rho_{ss}(P_k)\right]\exp\left(-k_{pol}(P_k)\cdot t_k\right), \quad (10)$$

where $k_{pol}(P_k)$ is the power-dependent effective polarization rate, defined as follows:

$$k_{pol}(P_k) = (C_1 + C_2)\alpha P_k + \frac{1}{T_1}, \quad (11)$$

where, $T_1$ is the longitudinal relaxation time of NV centers.

Compared with the traditional forward Euler method, the analytical discretization scheme in Eq. (10) provides the following advantages:

1) Elimination of stability constraints: Since $k_{pol}$ can reach the order of $10^8$ $s^{-1}$ under high laser power, traditional linear discretization demands a sampling interval far smaller than the characteristic evolution timescale. By contrast, the analytical mapping allows the agent to directly predict state transitions over a wider range of pulse durations (100 ns to 1600 ns) in the action space.

2) Physical consistency: This scheme fully preserves the nonlinear polarization dynamics within a single cycle and avoids cumulative numerical errors. This enables the agent to accurately capture the trade-off between polarization gain and the penalties imposed by laser power and pulse duration.

3) Training efficiency: Within the discrete training environment, the agent observes the state vector $s_k = [P_k, t_k]$ and explores the action space to identify the optimal control strategy. The polarization rate $\rho_{k+1}$ from each state transition is used to compute the reward $R$, which in turn drives the $Q$-table toward the physically global optimum.

## Ⅲ. Low-Power NV-Centers Polarization Efficiency Optimization Architecture

This section presents a detailed description of the RL framework designed to optimize the spin polarization efficiency of NV centers under minimum heating effect constraints. The objective of this framework is to autonomously identify the optimal trade-off between laser power $P$ and pumping duration $t$ while satisfying heating effect constraints. Fig. 2(a) illustrates the overall workflow of the offline training process, which is based on the analytical five-level dynamical model of the NV centers established in Section II. The agent consists of two core components: the actor and the learner. The actor interacts with the simulation environment and generates candidate laser power $P$ and pulse duration $t$ within the predefined action space, whereas the learner updates the strategy using the Bellman equation based on the polarization rate gains from environmental feedback and thermal safety constraint evaluation data. This offline training mode ensures that the control strategy can cover nanosecond-level time scales, while avoiding potential thermal damage to actual experimental equipment during initial exploration. Fig. 2(b) depicts a simulation environment based on the analytical discretization model of Eq. (12). This environment integrates a laser excitation module, an ISC dynamics module and a thermal constraint model. The simulation system can predict the heating effect under different control sequences in real time by incorporating the experimentally calibrated thermal-constraint coefficient into the environment, thereby providing physically consistent reward feedback to the agent. Fig. 2(c) illustrates the online deployment process of the strategy. The system derives an optimal closed-loop control strategy, which is subsequently integrated into the NV centers control platform. In practical ODMR experiments, the system employs the optimized laser pulse sequence to achieve a dynamic balance between spin polarization rate and thermal stability. Fig. 2(d) presents the core configuration of the NV centers experimental platform. The experimental setup consists of a confocal optical path, a microwave antenna, and a fluorescence collection module. An acousto-optic modulator (AOM) is utilized to modulate laser pulse over timescales ranging from nanoseconds to milliseconds. The collected fluorescence signals are detected by a single-photon counter and subsequently fed back to the control unit for real-time regulation.

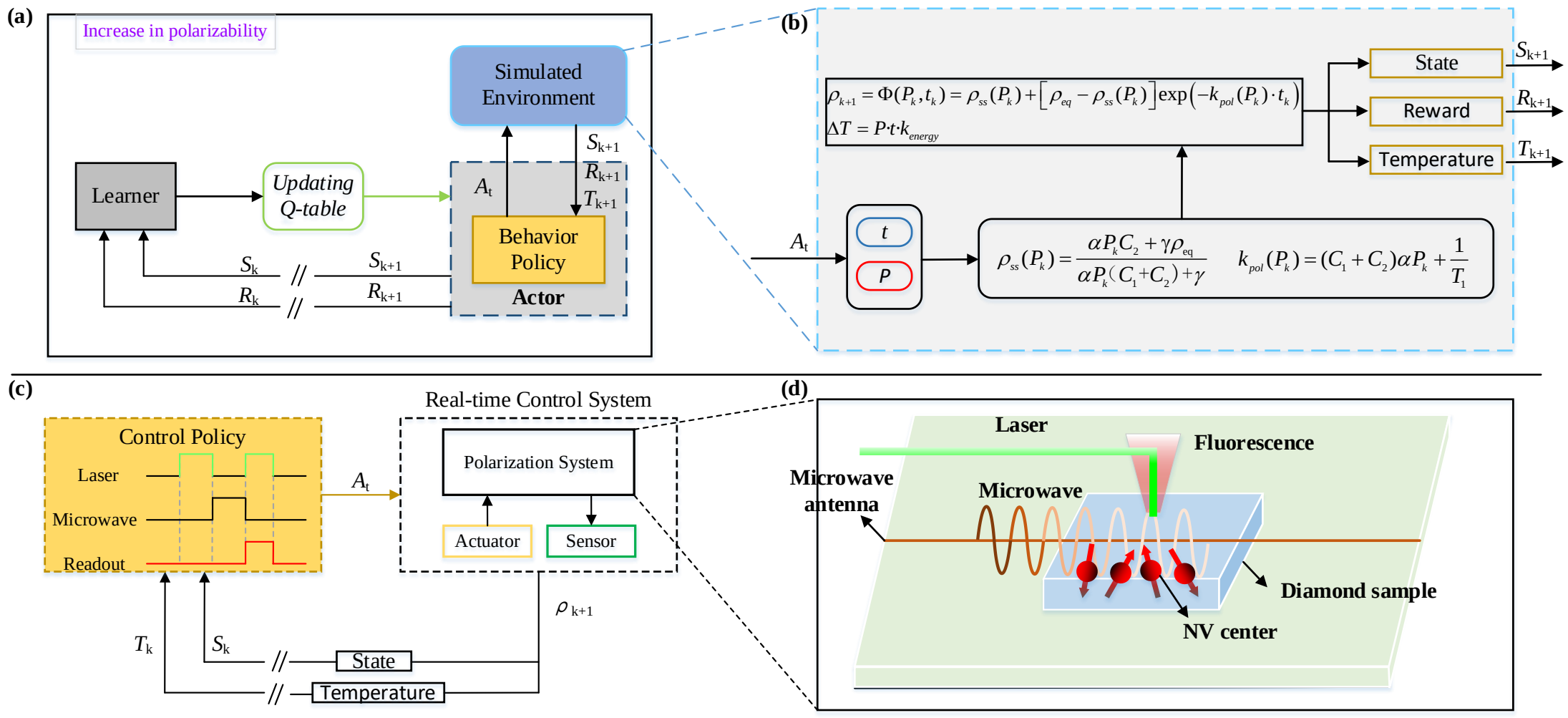

**Fig.1 Proposed closed-loop controller design architecture for NV centers spin polarization.**

**(a) Training loop: The offline optimization framework where the Learner updates the action-value space (*Q*-table) to guide the Actor's behavior policy based on state ($S_k$) and reward ($R_k$) feedback. (b) Simulated environment: The physics-informed simulation model that ingests control actions $A_t = (t, P)$ to synchronously solve the analytical steady-state polarization $\rho_{ss}$ ($P_k$), dynamic spin evolution $\rho_{k+1}$, and laser-induced thermal accumulation ΔT. (c) Online deployment: Real-time execution of the optimal control policy, translating the learned policy into precise timing sequences (Laser, Microwave, and Readout pulses) managed by the hardware actuators and sensors. (d) NV centers polarization system: Schematic of the physical experimental setup, including the diamond sample embedded with NV centers, microwave antenna for spin manipulation, and green laser excitation for initialization and fluorescence readout.**

## A. Establishment of MDP framework

RL enables agents to learn behavioral strategies that maximize the expected cumulative reward through continuous interaction with the environment. As a classic off-policy temporal difference algorithm, $Q$-learning enables the agent to collect experience through interaction with the environment and iteratively update the Q-table to optimize strategy[19]. To achieve the optimization goal, we formulate the spin polarization control process as a Markov Decision Process (MDP), whose core elements are defined as follows.

1) State Space: The state vector is defined as $s_k=[P_k, t_k]\in\boldsymbol{S}$, where $P_k$ and $t_k$ denote the laser power and pulse duration at the current step, respectively. For computational convenience, the power range of [0.1, 15.0] mW and the time range of [100, 1600] ns are uniformly discretized into $N$=40 intervals.

2) Action Space: The action set of the intelligent agent is defined as A={Up, Down, Left, Right}, which corresponds to the one-step increase/decrease adjustments to $P$ or $t$ in a discrete grid.

3) State Transition: The system evolves according to the analytical dynamical model. When action $a_k$ is applied, the state transitions from $s_k$ to $s_{k+1}$ and the corresponding spin polarization $\rho_{k+1}$ is calculated based on the analytical solution.

4) Reward Function: To balance polarization gain against physical resource consumption, the reward $R$ is defined as a multi-objective function consisting of a gain term, an energy penalty, and a thermal safety constraint.

$$R = r_{gain} - r_{power} - r_{time} - r_{thermal} \tag{12}$$

Polarization gain ($r_{\text{gain}}$): The logarithmic function $r_{\text{gain}}=\omega_1 \log(\rho+\phi)$ models the diminishing marginal returns of polarization gains.

Energy consumption penalty ($r_{\text{power}}$, $r_{\text{time}}$): The output $t$ by the algorithm is defined as the pumping duration. The heating effect $\Delta T$ is proportional to $t$. This relation constitutes the primary physical constraint of the PAS-Q algorithm. Exponential and linear penalties are applied to $P$ and $t$, respectively, to suppress excessive power and duration.

Thermal constraint ($r_{\text{thermal}}$): If the estimated heating effect $\Delta T= P \cdot t \cdot k_{\text{energy}}$ exceeds the threshold $T_{th}$, a large penalty is imposed to discourage the agent from entering high-temperature regimes.

$$r_{thermal} = \begin{cases} \Omega, & \Delta T > T_{th} \\ \omega_2 (\Delta T - T_{th})^2, & T_{th} < \Delta T \le T_{safe} \\ 0, & \Delta T \le T_{th} \end{cases} \tag{13}$$

## B. Minimum Heating Effect Constraint Model

In P-ODMR, the laser pulse injects energy that induces localized heating of the NV centers. Because the integration time (on the order of milliseconds) far exceeds both the single-pulse sequence period and the phonon thermal relaxation time, thermal accumulation results in a steady-state temperature distribution.

The effective duty cycle $\eta_{\mathrm{eff}}$ of the measured sequence is defined as:

$$\eta_{\mathrm{eff}} = \frac{t_{pol} + t_{ref} + t_{read}}{T_{sequence}}, \tag{14}$$

where $t_{\mathrm{pol}}$ is the polarization pulse width, $t_{\mathrm{read}}$ is the signal readout pulse width, $t_{\mathrm{ref}}$ is the reference signal pulse width, and $T_{\mathrm{sequence}}$ is the total duration of a single sequence.

A linear response model is established between the steady-state heating effect $\Delta T_{\mathrm{ss}}$ and the average laser power:

$$\Delta T_{ss} = k_{\mathrm{energy}} \cdot (P \cdot \eta_{\mathrm{eff}}), \tag{15}$$

where $P$ represents the peak laser power, and $k_{\mathrm{energy}}$ represents the laser thermal constraint coefficient to be calibrated. This model accounts for the integral energy of all laser-on periods in the sequence, providing an accurate estimate of the heating effect.

To determine $k_{\mathrm{energy}}$ under experimental conditions, a calibration method based on NV centers self-sensing is employed. The coefficient $k_{\mathrm{energy}}$ is calibrated by monitoring the frequency shift of the zero-field splitting parameter $D$ as a function of laser power. The experimental procedure is as follows:

1) The laser power $P$ is increased (from 5 mW to 35 mW) or the pulse duty cycle is adjusted. At each point, the system is held sufficiently long to allow the diamond lattice to reach thermal equilibrium with its surroundings.

2) The zero-field splitting energy $D$ is scanned and extracted in real time via a P-ODMR sequence. Laboratory temperature fluctuations are strictly controlled to ensure that any observed frequency drift arises solely from laser-induced heating. The experimental parameters are shown in Table 3.

3) The measured frequency $D$ is converted into the sample temperature $T$ using the known $D$-$T$ relationship for NV centers.

4) A linear regression model is constructed relating $T$ to the average laser power $P_{avg} = P \cdot \eta_{eff}$. The slope obtained by linear fitting gives the laser thermal constraint coefficient $k_{energy_}P$ = 0.0573 K/mW as shown in Figure 2. Table 4 shows the experimental protocol for varying polari zation time.

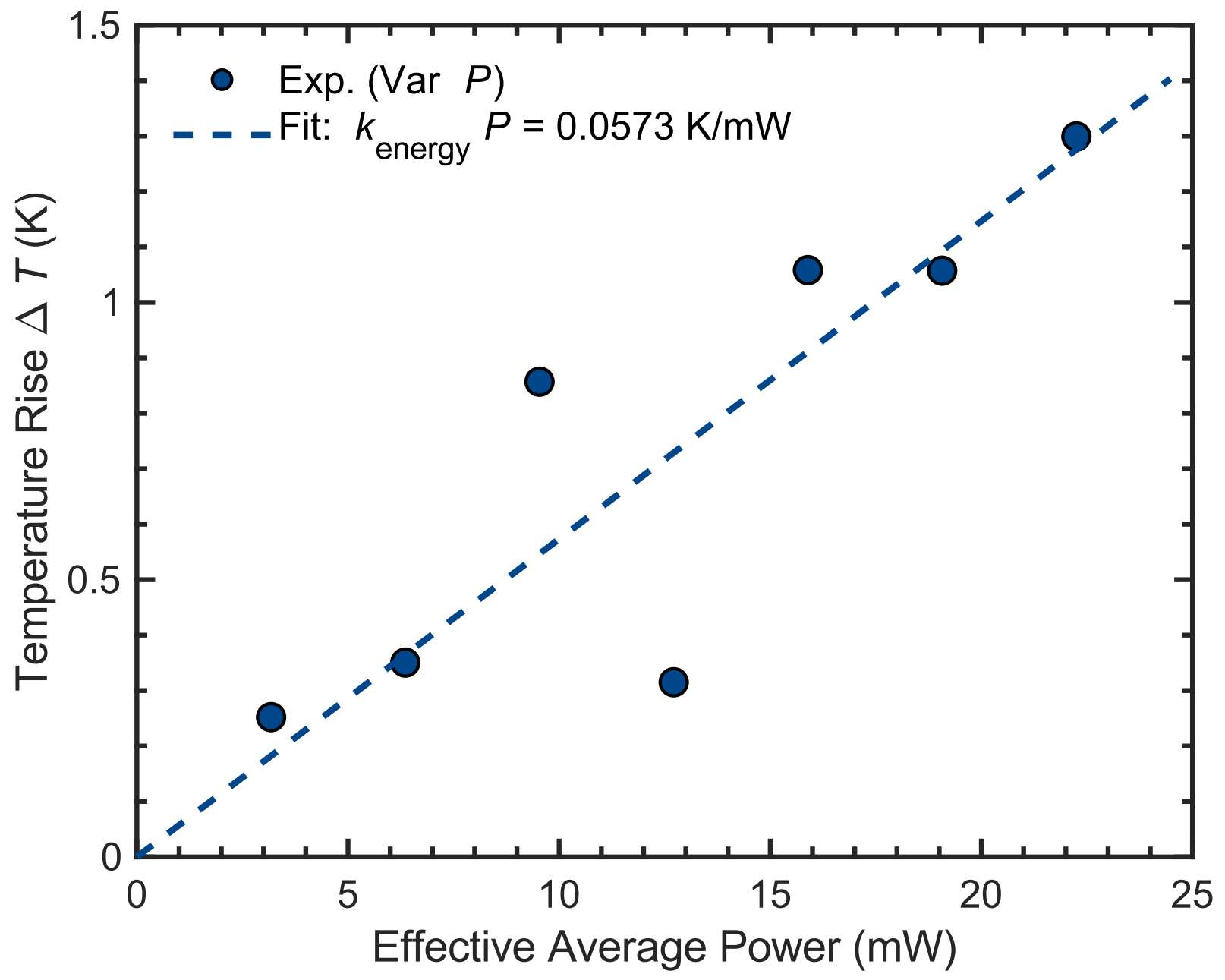

**Fig. 2 Thermal constraint coefficient of laser power**

**TABLE 1 Experimental Protocol for Measuring the Thermal Coefficient of Laser Power**

| Group | Laser power $P$(mW) | Polarization time $t_{pol}$(ns) | period of the sequence $T_{seq}$(ns) | ZFS $D$ (MHz) | Rising temperatures Δ$T$(K) |
|---|---|---|---|---|---|
| 1 | 5 | 1000 | 3907.6 | 2870.569046 | 0.2521 |
| 2 | 10 | 1000 | 3907.6 | 2870.561854 | 0.3505 |
| 3 | 15 | 1000 | 3907.6 | 2870.524795 | 0.8570 |
| 4 | 20 | 1000 | 3907.6 | 2870.564447 | 0.3150 |
| 5 | 25 | 1000 | 3907.6 | 2870.510043 | 1.0583 |
| 6 | 30 | 1000 | 3907.6 | 2870.510119 | 1.0573 |
| 7 | 35 | 1000 | 3907.6 | 2870.492355 | 1.2993 |

**TABLE 4 Experimental Protocol for Pulse Width Thermal Coefficient**

| Group | Laser power $P$(mW) | Polarization time $t_{pol}$(ns) | period of the sequence $T_{seq}$(ns) | ZFS $D$ (MHz) | Rising temperatures $\Delta T$(K) |
|---|---|---|---|---|---|
| 1 | 20 | 200 | 3107.6 | 2870.545345 | 0.576315 |
| 2 | 20 | 400 | 3307.6 | 2870.541216 | 0.632743 |
| 3 | 20 | 600 | 3507.6 | 2870.542118 | 0.620412 |
| 4 | 20 | 800 | 3707.6 | 2870.540494 | 0.642607 |
| 5 | 20 | 1000 | 3907.6 | 2870.527555 | 0.819348 |
| 6 | 20 | 1200 | 4107.6 | 2870.532000 | 0.758653 |
| 7 | 20 | 1400 | 4307.6 | 2870.534429 | 0.725474 |
| 8 | 20 | 1600 | 4507.6 | 2870.532734 | 0.748621 |
| 9 | 20 | 1800 | 4707.6 | 2870.530288 | 0.782021 |
| 10 | 20 | 2000 | 4907.6 | 2870.521240 | 0.905542 |

Based on the acquired real-time temperature measurement data, a regression model is constructed between the temperature $T$ and the average laser power $P_{avg} = P \cdot \eta_{eff}$. This allowed us to determine the laser thermal constraint coefficient at the pulse width, $k_{energy}$_$T$=0.0567 K/mW, as shown in Figure 3.

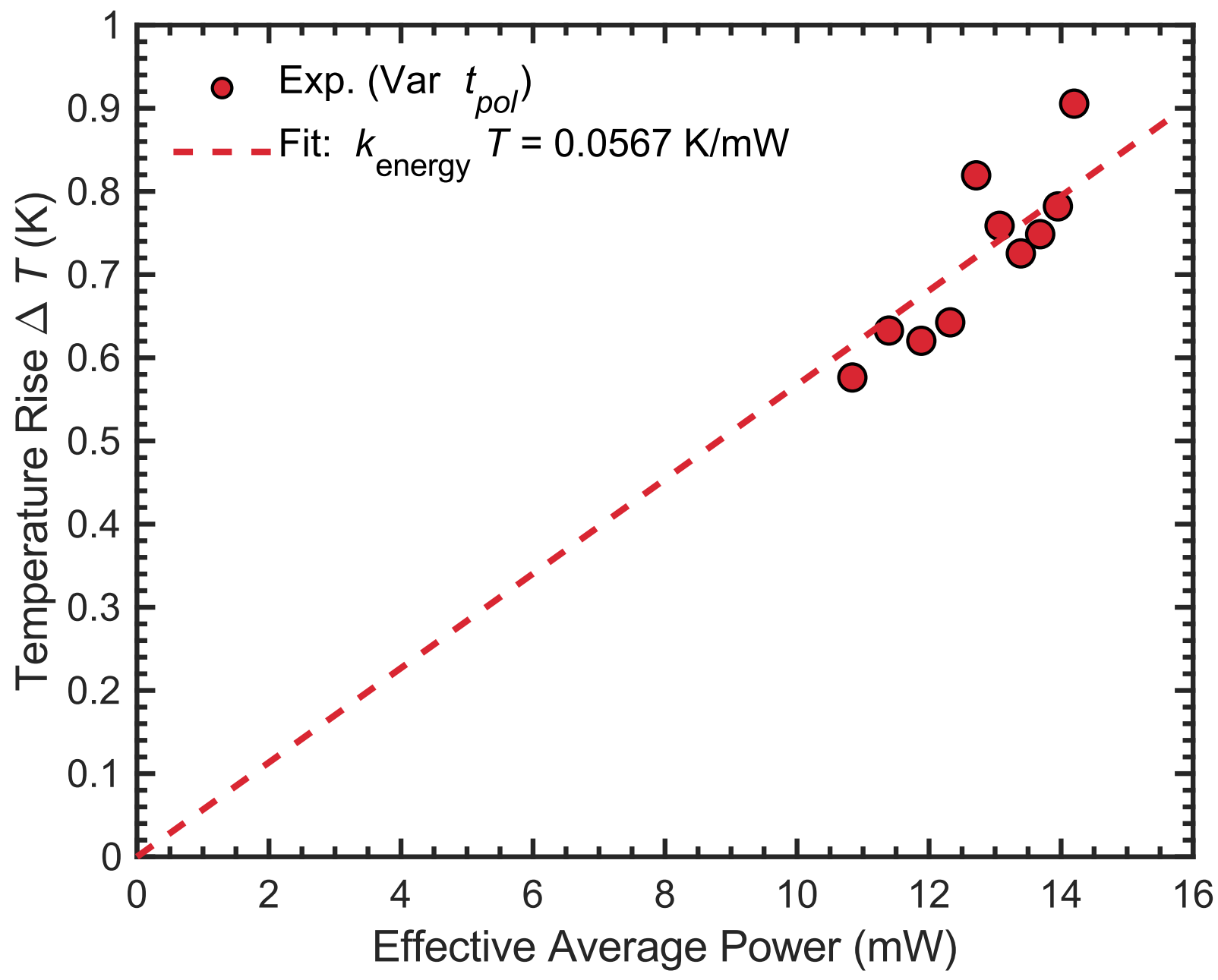

**Fig. 2 Thermal constraint coefficient of pulse width**

Finally, two independent data based on optical power and pulse width are fitted to eliminate potential systematic biases from a single experimental dimension and obtain a universally applicable thermal constraint model, as shown in Figure 4. The final laser thermal constraint coefficient is determined to be $k_{energy}$=0.0570 K/mW, which constrains the subsequent algorithm model.

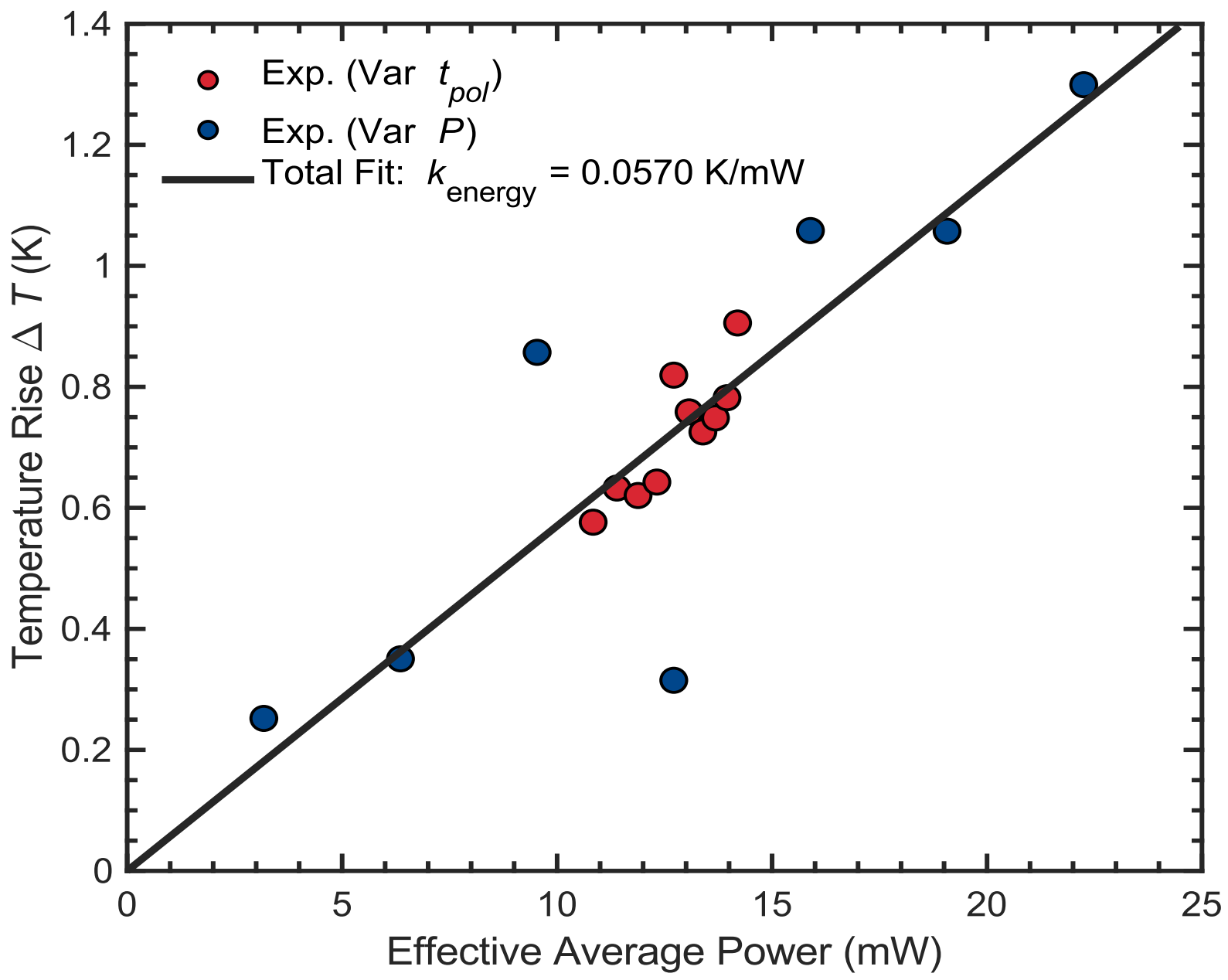

**Fig. 3 Fitting results of thermal constraint coefficient**

We propose a novel PAS-Q algorithm that incorporates dynamic action-space constraints and physical thermal limits. The algorithm features a progressive constraint-relaxation mechanism that gradually lifts the laser power limit as training proceeds. This mechanism gradually removes the power limit as training progresses, guiding the agent from a low-power safe zone toward a high-polarization challenge zone. In this way, the problem of slow convergence caused by frequent exploration in the high-temperature penalty zone at the beginning of training is effectively avoided. Consequently, the polarization rate exhibits a distinct stepwise evolution during training and ultimately stabilizes at an optimal steady state satisfying the thermal constraints.

**Algorithm 1** PAS-Q Algorithm

**Input:** Initialize parameters according to Table.II;

**Output:** Optimal strategy $\pi$*and optimal parameters ($P$*, $t$*)

Initialize action-value function Q($s$, $a$) ← 0 for all s ∈ S, a ∈ A

Initialize historical best polarization $\rho_{best} \leftarrow 0$

**for** episode = 1, 2, ..., $N_{episodes}$ do

Compute dynamic power boundary $P_{max}$ to enable curriculum learning

Initialize starting state $s_0 = (P_0, t_0)$ randomly within valid boundary

**for** step = 1, 2, ..., N_steps do

With probability ε select a random action a ∈ A

Otherwise select a = $\mathrm{argmax}_{a'}$ $Q$(s, a')

Execute action a, observe candidate next state $s_{next} = (P', t')$

**if** $s_{next}$ exceeds boundary or $P' > P_{max}^{(ep)}$ then

Receive boundary penalty $r = -20$, keep state $s_{next} = s$

**else**

Calculate polarization $\rho_{00}$ using (11)

Calculate average thermal load $P_{avg} = P \cdot \eta_{eff}$

Predict steady-state temperature rise using (15)

Get the $R_{t+1}$ using (13)

**if** $\rho_{00} > \rho_{best}$ then Update optimal parameters $(P^*, t^*) \leftarrow (P', t')$

**end if**

$Q(s, a) \leftarrow Q(s, a) + \alpha \times [r + \gamma \times \max_{a'} Q(s_{next}, a') - Q(s, a)]$

State transition $s \leftarrow s_{next}$

**end for**

Update exploration rate: $\varepsilon \leftarrow \max(\varepsilon \times \varepsilon_{decay}, \varepsilon_{min})$

**end for**

**Return** $(P^*, t^*)$

The PAS-Q algorithm integrates the ISC analytical dynamics with the thermal constraint model. It directly outputs the spin polarization $\rho_{00}$ and the estimated heating effect $\Delta T$ at each state transition, thereby supplying precise physical gradients for policy optimization. Based on this, a multi-objective reward function is formulated with four mutually balancing components. A logarithmic gain term $r_{gain}$ that captures the diminishing marginal returns of polarization improvement. Exponential and linear penalty terms $r_{power}$ and $r_{time}$ that constrain peak power and time cost, respectively. A segmented thermal constraint $r_{thermal}$ that imposes a quadratic penalty when the temperature rise exceeds the safety threshold $\Delta T$. This guides intelligent agents to steadily expand from low thermal safety zones to high-dimensional spaces by dynamically relaxing power boundaries, ultimately converge to a globally optimal control strategy ($P^*$, $t^*$) that achieves high spin polarization with minimal laser-induced heating.

# Ⅳ. Simulation Results and Discussion

## A. Agent training and results

This section verifies the effectiveness of the proposed PAS-Q algorithm in NV centers polarization optimization through numerical simulation. The simulation environment is based on the analytical dynamic model established in Section II, focusing on investigating the algorithm's optimization trajectory and convergence performance under multiple constraints.

Figure 5 illustrates the convergence curves of the maximum polarization rate, exploration rate, and reward value during the 20,000-round training process of the PAS-Q algorithm. At the beginning of training, the exploration rate begins to decay exponentially, prompting the agent to conduct extensive random exploration of the action space in the early stages. During this period, the cumulative reward value exhibits a sharp upward trend as the agent continuously makes trial and error and updates its experience. Simultaneously, the maximum spin polarization rate of the NV centers also increases rapidly from its initial value. The exploration rate drops below 0.2 as training progresses to 6,000 rounds and continues to decay gradually towards 0, indicating that the algorithm enters a deep development state. During this phase, the maximum polarization rate curve stabilizes at an optimal level. Simultaneously, the reward value curve stops rising and gradually converges to a range between $2.3\times10^4$ to $2.4\times10^4$, with the corresponding variance shadow area significantly narrowing.

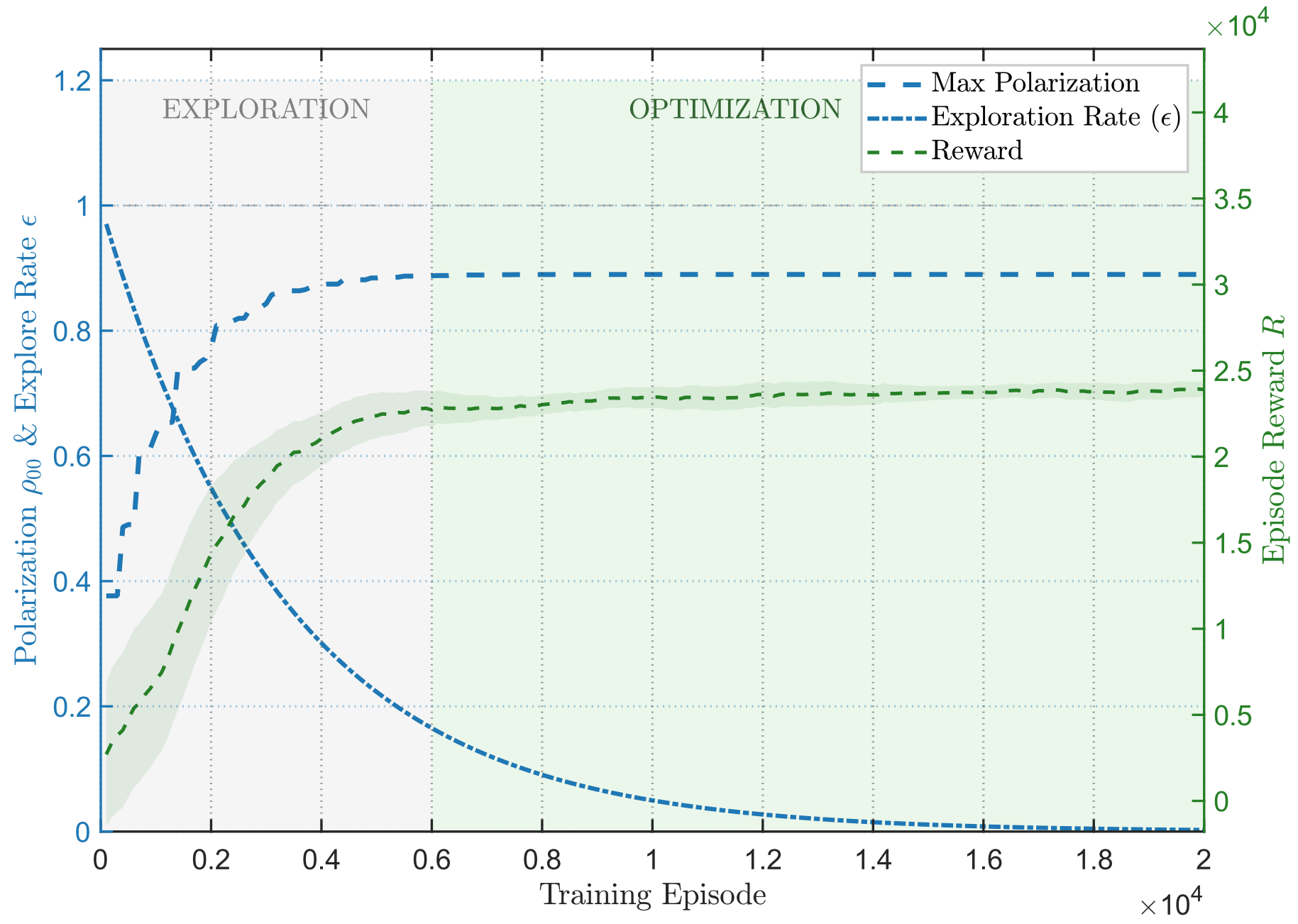

**Fig. 4 Convergence dynamics of the reinforcement learning training process. Evolution of the maximum spin polarization $\rho_{00}$ (blue dashed line) and the exploration rate ε (blue dash-dotted line) on the left axis, alongside the episodic reward $R$ (green dashed line, right axis). The shaded green region represents the variance of the reward across the training progress. The background colors distinguish two primary training phases: the initial 'EXPLORATION' phase, where a high decaying ε allows the agent to extensively search the action space, and the subsequent 'OPTIMIZATION' phase, where the agent exploits the learned policy.**

The optimization trajectory of the agent on the two-dimensional reward landscape is shown in Fig. 6. The colored contour map represents the scalar distribution of the reward function $R$, with high reward areas concentrated in the combination region of moderate power and long pulse duration; penalty areas correspond to extremely low polarization regions or high energy consumption regions exceeding temperature rise constraints.

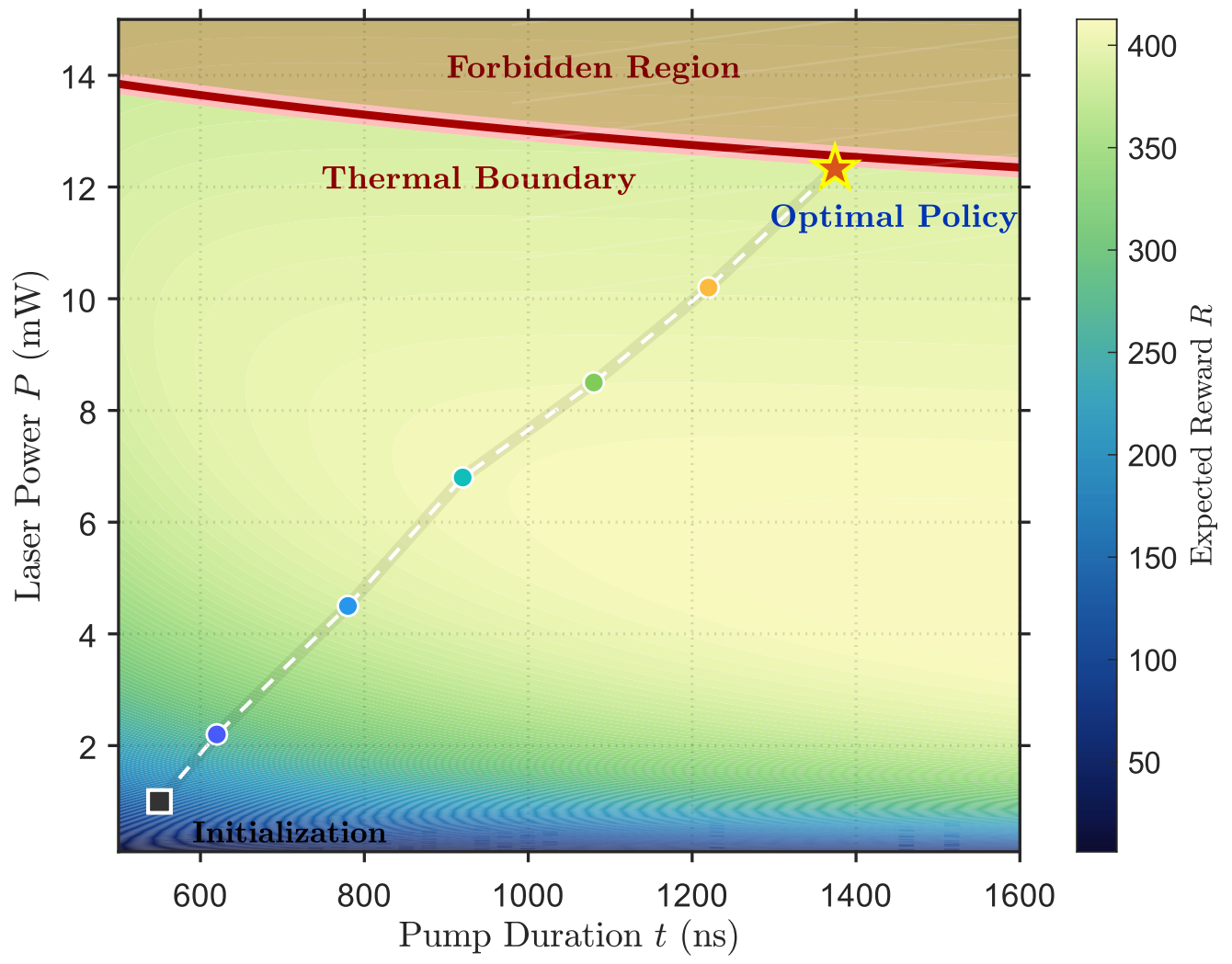

**Fig. 5 Optimization trajectory of the PAS-Q algorithm within the reward topology space. The dashed white line traces the evolution of the agent's policy from the initial exploratory state (black square) through intermediate updates (colored circles) to the optimal policy (yellow star). The solid red line demarcates the strict physical thermal boundary, above which lies the forbidden region associated with excessive laser-induced heating.**

The white dashed line represents the training trajectory of the agent, starting from the initial state (low power, short pulse) and successfully avoiding the Thermal Limit above, ultimately converging to the optimal operating point marked by the red pentagram at $P$=12.33 mW and $t$=1374.36 ns. This indicates that the RL controller is capable of autonomously learning the balance between excitation rate and heating effect.

## B. Polarization Dynamics

Using optimized control parameters combined with Eq. (11), we simulate the evolution process of NV centers from thermal equilibrium to steady state, as shown in Fig. 7. When reaching the same polarization ratio $\rho$ = 0.80, the suboptimal control schemes of 2.5 mW and 7.4 mW consumed approximately 1000 ns and 350 ns, respectively. However, the optimized control strategy shortens this process to approximately 200 ns and ultimately stabilizes at the optimal polarization $\rho_{00}$ = 0.890 at $t$=1374 ns. Although the suboptimal scheme of 14.8 mW can reach the same polarization rate within about 200 ns, its average heating effect increases significantly. Furthermore, the analytical discretization method ensures that the calculation does not produce numerical divergence, thereby providing the control strategy with extremely high robustness on the nanosecond scale.

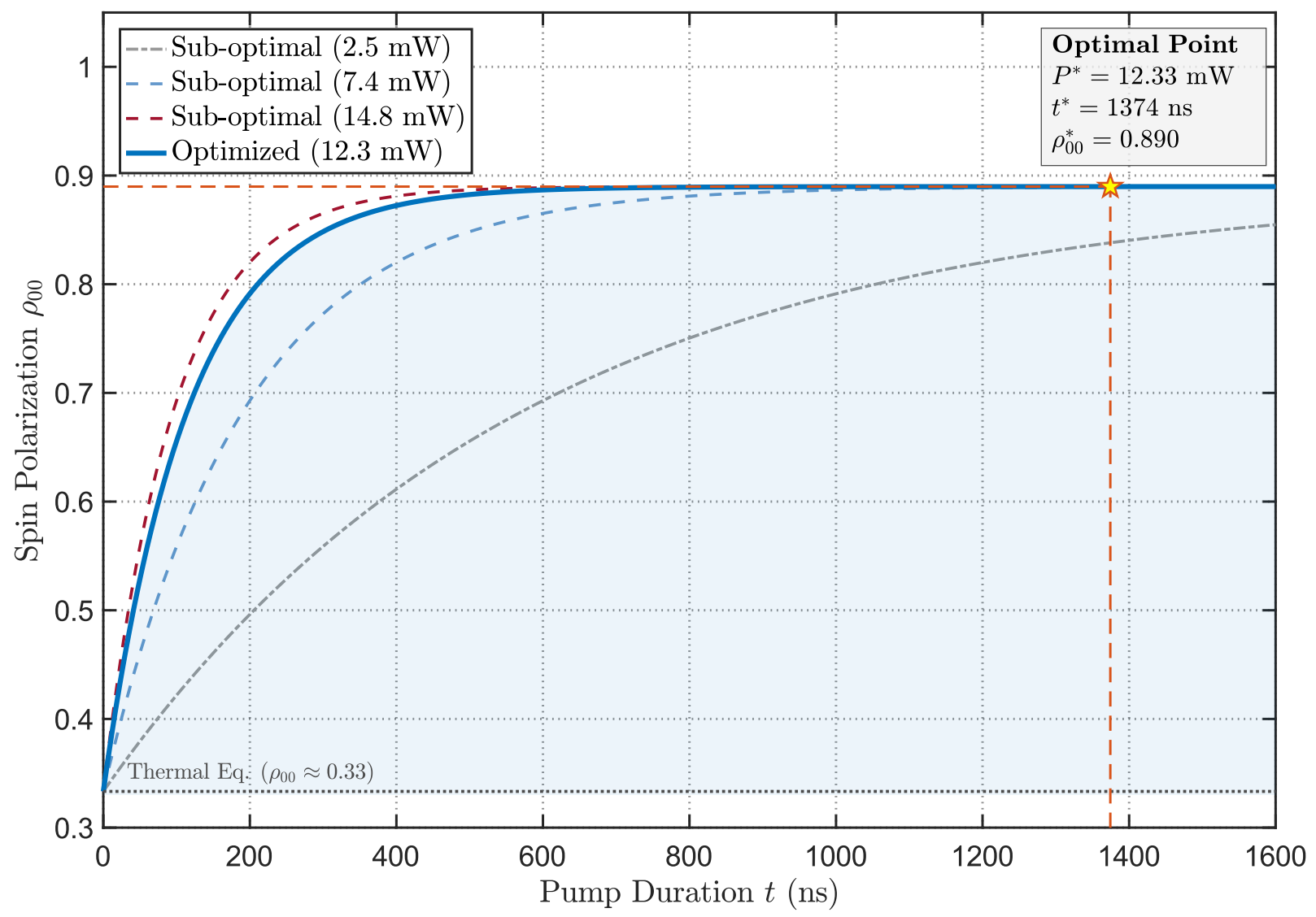

**Fig. 6 Optimization of spin polarization dynamics in NV centers. The solid blue line represents the optimized trajectory ($P$ = 12.3 mW), which is evaluated against three sub-optimal baselines: 2.5 mW (grey dash-dotted line, showing insufficient polarization rate), 7.4 mW (light-blue dashed line), and 14.8 mW (dark-red dashed line). The dotted horizontal line denotes the initial thermal equilibrium state ($\rho_{00} \approx 0.33$). The orange star explicitly marks the derived optimal operational point ($P^*$ = 12.33 mW, $t^*$ = 1374 ns, $\rho_{00}$ = 0.890), which successfully balances the initialization efficiency and laser-induced thermal constraints.**

# Ⅴ. Online deployment and analysis

## A. Experimental environment and deployment

Figure 8 depicts the experimental system constructed to verify the PAS-Q algorithm in a physical system. The platform mainly consists of three modules: algorithm decision-making, hardware execution, and signal feedback. The PAS-Q optimal $Q$-table, obtained from preliminary training, is stored as a discrete lookup table and deployed on the computer. During the experimental operation phase, the host computer directly calls the optimal state vector $s=[P^*, t^*]$, outputting it and converting it into a hardware control instruction. After receiving the instruction, the system adjusts the laser output power $P$ . The pulse generator then triggers the RF driver to control the AOM, timing, thereby regulating the laser pulse width $t$. A laser focused by a confocal optical system excites the NV centers, which then emit a fluorescence signal. After filtering, a single photon counter (SPCM) collects the fluorescence signal, a data acquisition card (DAQ) digitizes it, and the result is transmitted back to the computer in real time.

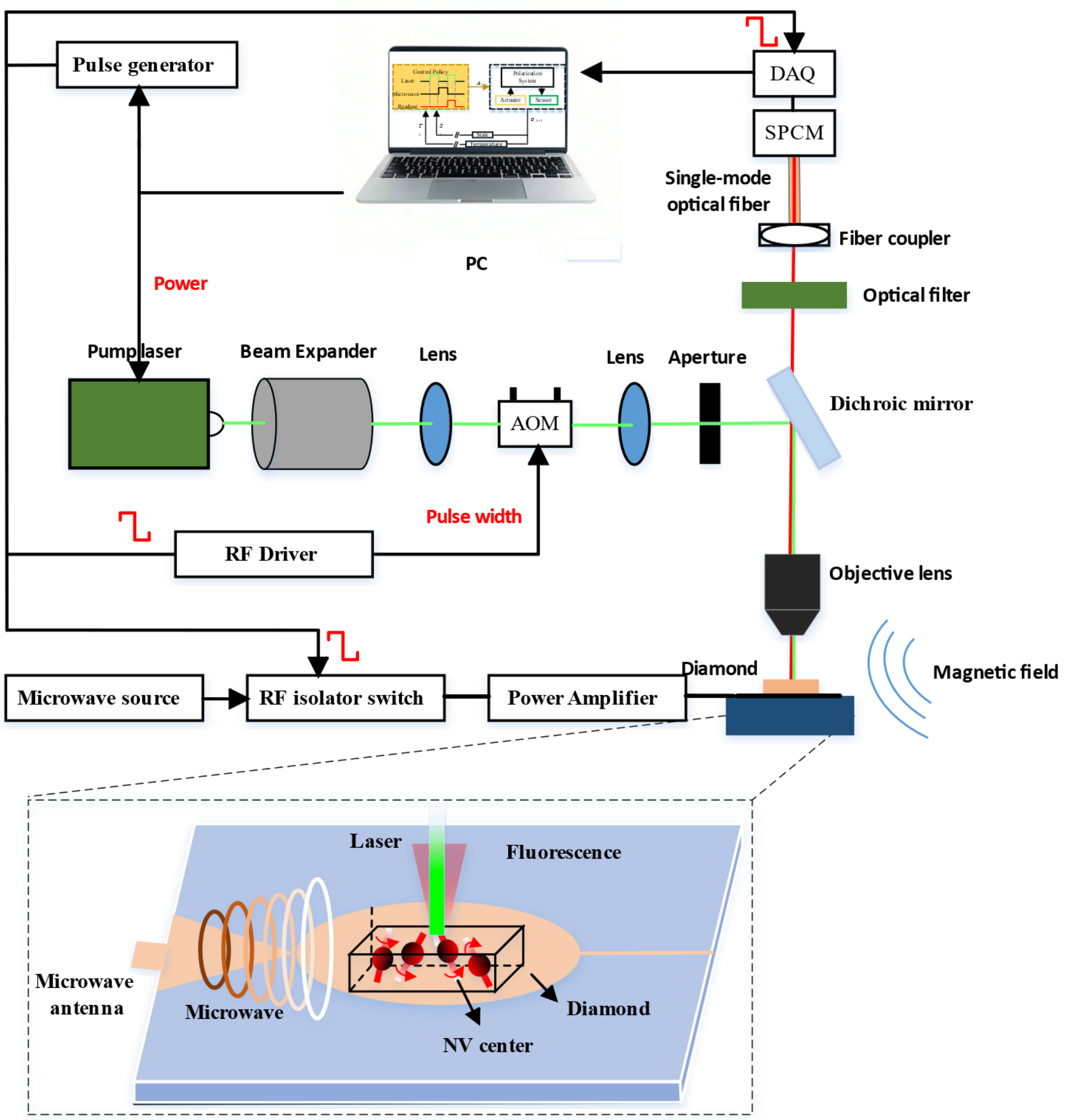

**Fig. 7 Experimental setup of the NV centers polarization.** SPCM: Single Photon Counting Module DAQ: Data acquisition system, PC: Personal computer. AOM: Acousto-optic modulator

Table 3 presents the optimal Q-values obtained by the intelligent agent after training. When the laser power is merely 5.00 mW, the system polarization rate is evidently inadequate. At this point, the *Q*-value is 215.42, which indicates that the agent perceives considerable improvement potential and makes a decision to increase the laser excitation intensity to escape the low-efficiency range. The system polarization is insufficient due to a short cycle time when the power rises 12.30 mW but the duration is only 500 ns. The agent outputs a Q-value of 289.12, which directs the system to prolong the polarization cycle for higher physical benefits. The system hits the predefined thermal constraint boundary when the power reaches 14.50 mW and the duration reaches 1500 ns, and the Q-value is negative. By contrast, the *Q*-value reaches its maximum of 414.19 at a power of 12.33 mW and a duration of 1374.36 ns. This discrete decision-making mechanism based on Q-tables realizes dynamic balance between polarization efficiency and thermal stability via predicting the gradient variation trend of Q-values on the nanosecond timescale.

**TABLE 1 *Q*-Table**

| State ($s_k$) | Power $P$ (mW) | time $t$ (ns) | $Q(s^*, a^*)$ | Strategy Interpretation |
|---|---|---|---|---|
| 1 | 5.00 | 800 | 215.42 | Insufficient polarization rate, guide to increase excitation intensity |
| 2 | 14.50 | 1500 | -58.10 | Trigger thermal constraints and guide back to the safe zone |
| 3 | 10.00 | 1100 | 328.65 | Approaching the physical equilibrium point |
| 4 | 12.33 | 1374.36 | 414.19 | Global *R*-optimal steady-state point |
| 5 | 12.30 | 500 | 289.12 | Sufficient power, guiding the extension of polarization cycle |

## B. Experimental Results and Analysis

This section compares the spin polarization efficiency and thermal constraints of the traditional empirical parameter scheme and the proposed PAS-Q optimization scheme. Since the spin-lattice relaxation of NV centers strongly relies on the phonon Raman process, heating effect can greatly accelerate electron nonradiative thermalization, thereby markedly shortening the $T_1$ time and fundamentally restricting the system-achievable polarizability. By comparing the $T_1$ decay curves obtained under different schemes, the interference of thermal constraints with spin polarization can be fully characterized. The initial amplitude at $\tau \to 0$ directly reflects the laser-induced polarization, while the $T_1$ time constant quantifies the impact of non-equilibrium thermal phonon scattering.

Traditional polarization parameter optimization is generally implemented under fixed laser power conditions. Stepwise pulse-width scanning and exponential fitting of fluorescence decay curves are used to determine the optimal polarization duration. However, this univariate control strategy cannot realize collaborative joint optimization of laser power and pulse width. Moreover, it relies excessively on redundant data sampling and post-fitting of full-range polarization curves.

For comparison, experiments are conducted at fixed laser powers of 10, 15, and 20 mW to evaluate the proposed algorithm. Figure 9 compares the $T_1$ decay dynamics measured under different parameter schemes. Intense laser-induced thermal effects shorten the $T_1$ value to merely 561.56 μs under the traditional 20 mW scheme. In contrast, the PAS-Q control strategy improves $T_1$ by approximately 49% relative to the traditional 20 mW scheme. Even when compared with the low-power 10 mW traditional scheme with $T_1$ = 629.56 μs, its $T_1$ value is still prolonged by around 33%. This result demonstrates that the proposed algorithm can effectively suppress high-temperature spin-phonon relaxation channels by precisely limiting the laser-induced thermal load. This alleviates the inherent trade-off between high spin polarization and laser-induced heating that exists in conventional strategies, thereby enabling global optimization of NV centers polarization with minimal heating.

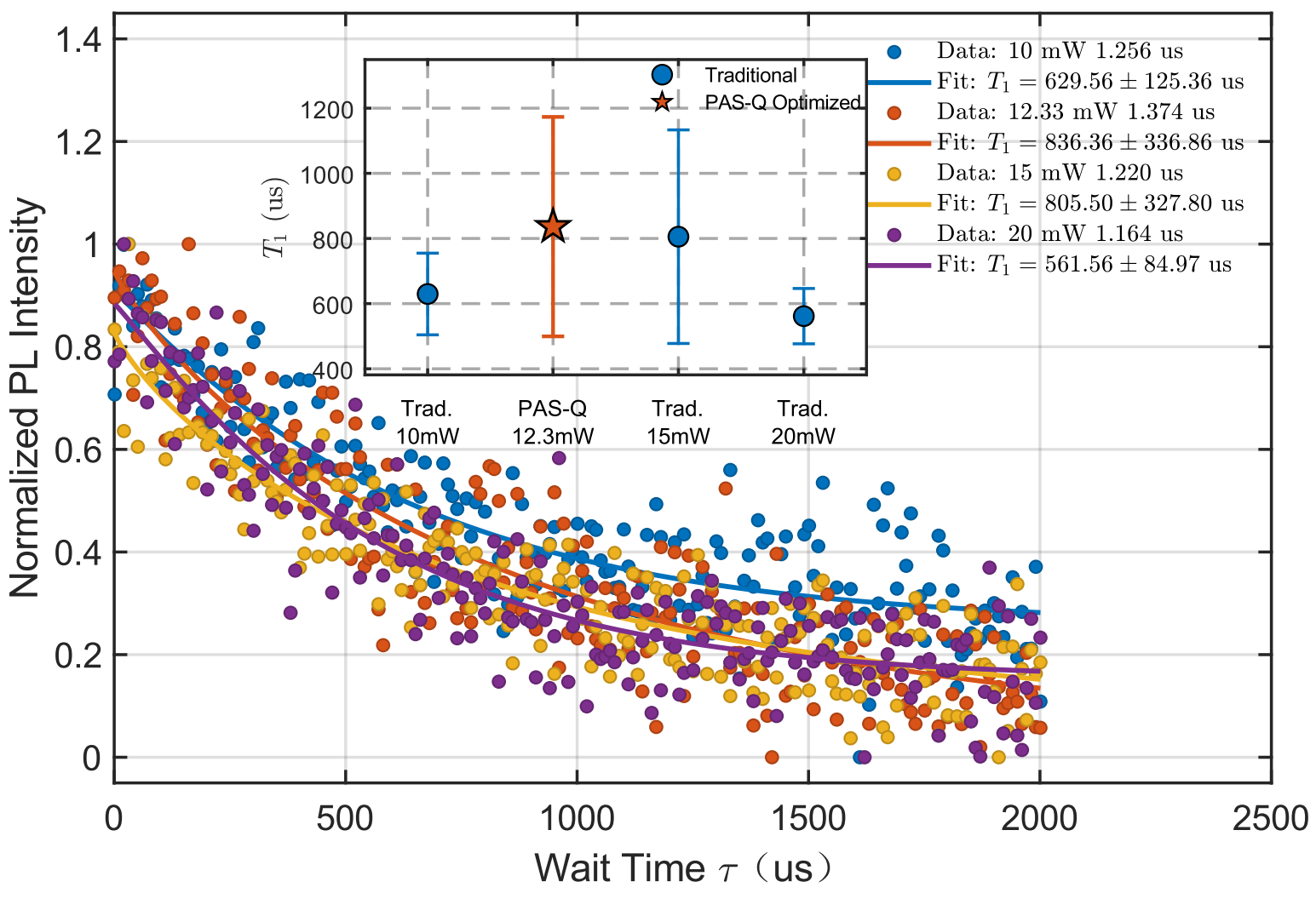

**Fig. 8 $T_1$ curves under different control strategies. Scatters represent experimental raw data, and solid lines indicate the corresponding single-exponential decay fits. Inset: Comparison of the extracted $T_1$ relaxation times with error bars denoting the fitting uncertainties.**

Furthermore, Figure 10 compares the computational time consumption of different optimization methods. At a sampling rate of 200 kHz, the traditional traversal-based parameter optimization scheme requires a total of 884.631 s. In comparison, the PAS-Q algorithm proposed herein converges toward the globally optimal control strategy within only 256 s via autonomous state-space exploration, improving the optimization efficiency by approximately 71%.

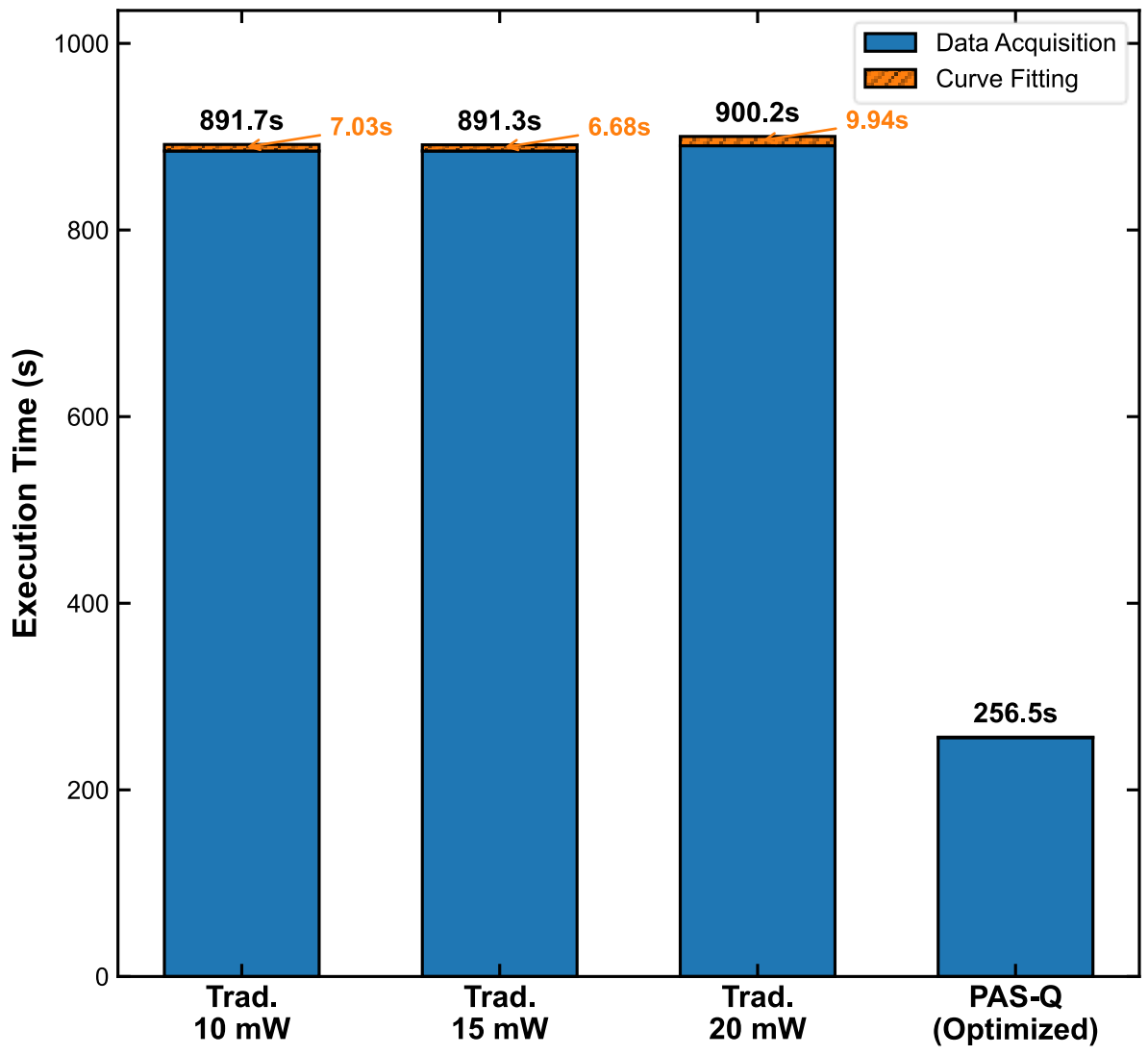

**Fig. 9 The time required for different control strategies to optimize polarization time**

P-ODMR spectroscopy is adopted to compare the control strategies derived from the three traditional univariate parameter optimization methods with the outputs of the proposed PAS-Q algorithm, so as to intuitively validate the algorithm's optimization performance. As illustrated in Fig.11, the resonance spectrum acquired under the PAS-Q control strategy achieves a spectral contrast of 4.11%. In comparison, the traditional schemes at laser powers of 10 mW, 15 mW, and 20 mW yield contrast values of 4.05%, 3.92%, and 3.65%, respectively. The proposed algorithm improves the spectral contrast by 1.48%, 4.85%, and 12.60% correspondingly, demonstrating its superior optimization performance.

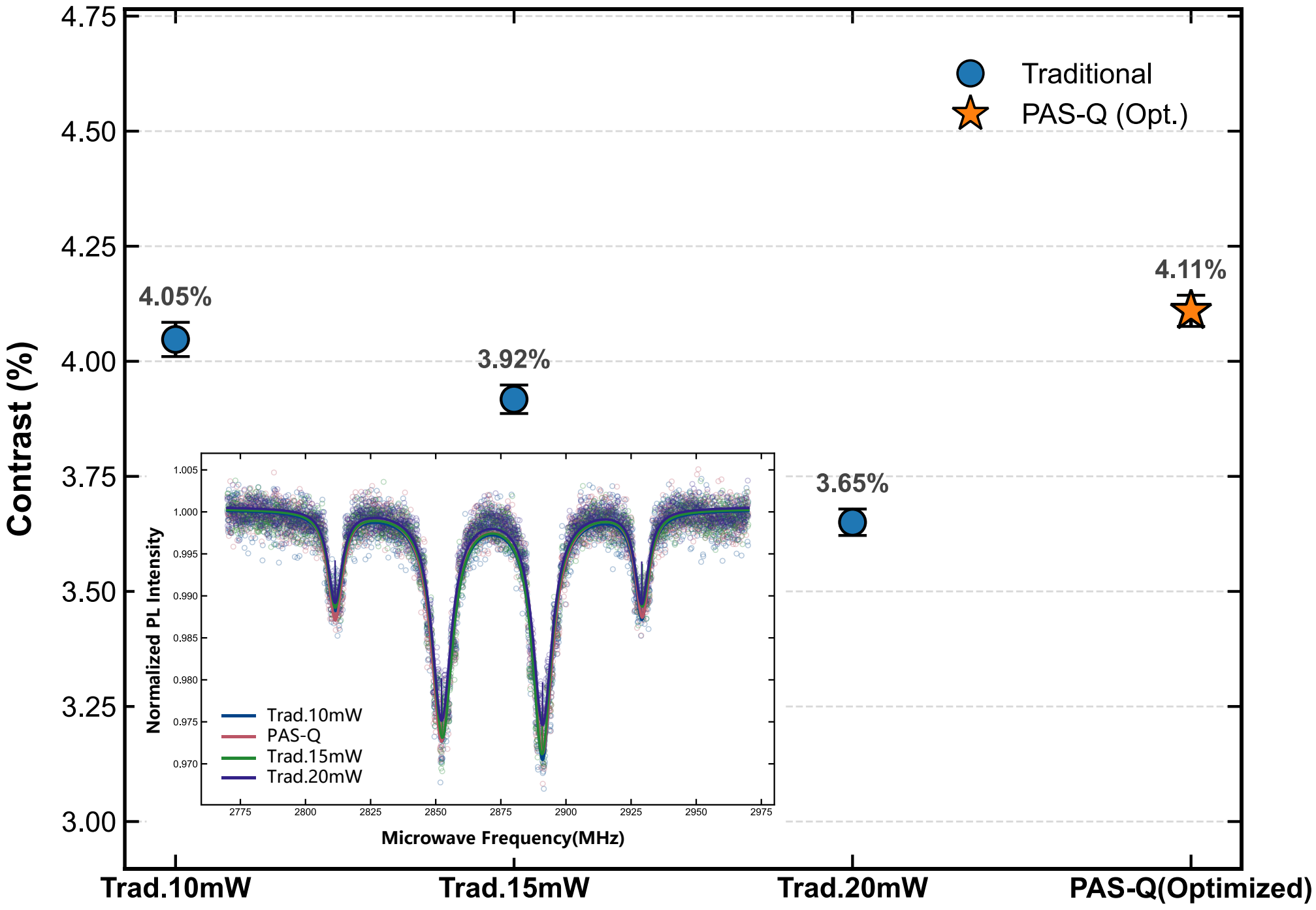

**Fig. 10 Contrast of ODMR spectroscopy under different power combinations. Comparison of the continuous-wave ODMR contrast extracted from the PAS-Q optimized control policy against three traditional constant-power baselines. Error bars represent the statistical uncertainties derived from the Lorentzian fitting of the resonances. Inset: The corresponding normalized photoluminescence intensity spectra as a function of microwave drive frequency, showing the characteristic NV center resonance dips.**

Figure 12 presents real-time monitoring data collected by synchronously arranged thermocouples. The PAS-Q scheme maintains an extremely low temperature rise amplitude throughout operation. By contrast, parameter combinations from traditional schemes cause experimental temperature rises fluctuating around 1 K. In comparison, the algorithm proposed herein achieves an experimental temperature rise of only 0.56 K, which is approximately 44% lower than that of traditional methods. This reveals that the PAS-Q algorithm fulfills the optimization targets of minimal heating effect and high spin polarization efficiency, while effectively reducing optimization time consumption.

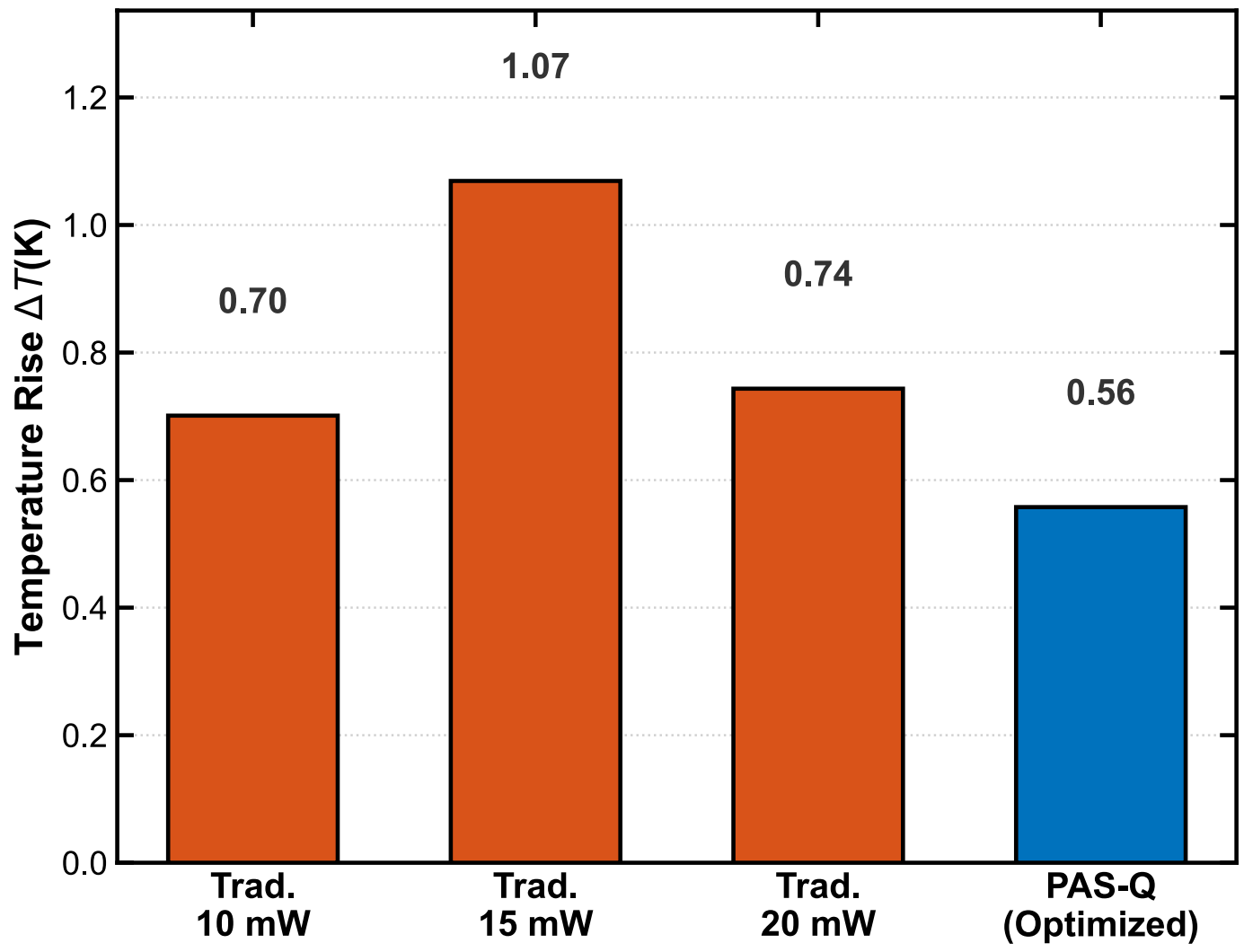

**Fig. 11 Heating effect under different power combinations**

# Ⅵ Conclusion

This paper proposes and experimentally validates a PAS-Q algorithm to optimize the spin polarization rate of NV centers under minimal temperature rise. The proposed method addresses the physical bottleneck of quantum sensing accuracy induced by laser thermal effects in P-ODMR systems. By incorporating analytical ISC dynamics and a steady-state thermodynamic framework, the algorithm enables efficient autonomous global optimization of spin polarization efficiency with minimal thermal generation. When deployed in the P-ODMR system, the PAS-Q-optimized control strategy yields a 33% improvement in polarization rate and a 71% enhancement in optimization efficiency compared with traditional empirical parameter-based strategies. Meanwhile, it reduces the laser-induced thermal load by 44% relative to traditional schemes. These findings facilitate the practical engineering deployment of NV centers quantum sensors. Moreover, this work introduces RL into the field of quantum instrument optimization, providing a novel technical paradigm. Nevertheless, the current approach only relies on steady-state characteristics for optimization. In our subsequent work, we will consider a collaborative optimization model of microwave and laser to further reduce the impact of heating effects.

## Funding

This work was supported in part by Fundamental Research Program of National

Institute of Metrology, China (No. AKYZD2604-2); National Natural Science Foundation of China (Grant 62505304); Exploratory Innovation Project of National Institute of Metrology, China (No. AKYCX2401); Natural Science Foundation of Zhejiang Province （Grant LQN26F030046）